# A self-healing tactile sensor using an optical waveguide

Seiichi Yamamoto[1], Hiroki Ishizuka[1], Sei Ikeda[1] and Osamu Oshiro[1]

[1] *Graduate School of Engineering Science, Osaka University, Toyonaka, Japan*

(Email: yama19991015@bpe.es.osaka-u.ac.jp)

**Abstract ---** We propose an optical tactile sensor using self-healing materials. The proposed tactile sensor consists of a structure that includes a diode, a phototransistor, and an optical waveguide made from self-healing materials. This design offers the advantage of being less susceptible to electromagnetic noise compared to traditional tactile sensors based on electrical detection principles. The sensor estimates the applied force by detecting changes in the total internal reflection caused by deformation due to contact force. In this study, we first established a fabrication method for the optical waveguide-based tactile sensor using self-healing materials. Subsequently, we measured the sensor output when a static load was applied to the fabricated tactile sensor and evaluated its characteristics. The results confirmed that the sensor output decreases in response to the applied load.

## 1 INTRODUCTION

Conventional robots typically consist of stiff materials such as metal and plastic and are primarily designed for simple tasks such as factorial automation. Recent advancements in robotics have been significant, expanding the potential applications of robots across various fields, including industrial, medical, and service fields [1].

In such cases, the robots must flexibly adapt their behavior to the surrounding environment. Then, the robots utilize visual and tactile information to recognize their state and the surrounding environment. To correct tactile information, it is essential to embed a tactile sensor in the robots. The tactile sensor can capture touch events, such as touching or grasping objects. Recently, soft tactile sensors utilizing elastomers or gels have actively developed with material technology development. The elastomers and gels offer excellent toughness and elasticity, making soft tactile sensors highly impact-resistant. However, they are vulnerable to damage from cuts or punctures.

To address this issue, soft tactile sensors using self-healing polymers that can recover from cuts and punctures have been developed. The detection principles of these sensors primarily include resistive sensing, where electric resistance is changed by external stimuli [2], and capacitive sensing, which utilizes structures that change their capacitance under applied force [3]. However, these detection methods are susceptible to electromagnetic noise, limiting their applicability in environments that require high reliability. In response to this limitation, Bai et al. recently proposed a self-healing strain sensor for shape estimation that utilizes changes in light intensity caused by the deformation of an optical waveguide [4]. Although this detection principle can be applied to self-healing tactile sensors, research on optical self-healing tactile sensors has never been investigated so far. Optical self-healing tactile sensors differ structurally from strain sensors. To ensure reliable force detection, it is necessary to establish the design of the tactile sensor. Therefore, we develop an optical tactile sensor using self-healing materials in this study. We establish a method for fabricating an optical waveguide-based tactile sensor using a self-healing polymer, a diode, and a phototransistor, and evaluate its characteristics.

## 2 PRINCIPLE

The proposed self-healing tactile sensor utilizes an optical waveguide structure. Fig. 1 shows the state of the optical waveguide-based tactile sensor before and after applying an external force. The optical waveguide consists of two layers: a core that serves as the light path and a cladding surrounding it. The area where the force is applied is composed of a black-colored self-healing polymer, which prevents external light from affecting the contact area. In this design, the core is made of a self-healing polymer, while the cladding is air. Under normal healing polymer, while the cladding is air. Under normal conditions, light from the source undergoes total internal reflection at the boundary due to the difference in refractive indices between the core and the cladding materials, transmitting the light to the detector (Fig. 1(a)).

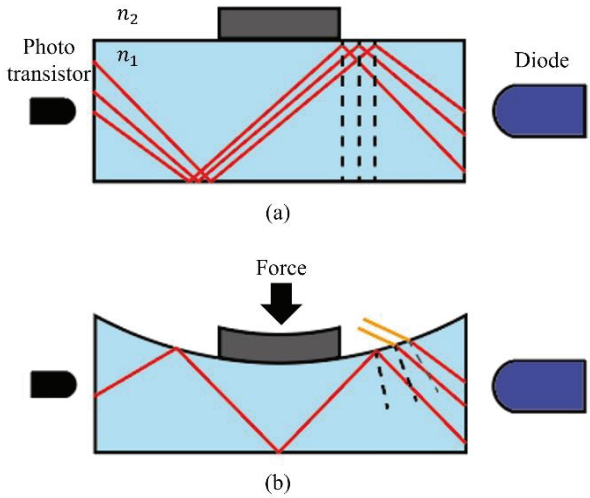

Fig.1 Principle of proposed sensor, (a) Before force is applied, (b) After force is applied.

The optical waveguide-based tactile sensor estimates the applied force by detecting changes in the amount of total internal reflection caused by the deformation of the optical waveguide due to external force (Fig. 1(b)). When force is applied to the optical waveguide, deformation occurs, altering the light path. As a result, the conditions for total internal reflection are disrupted, and light that enters at angles below the critical angle is scattered, leading to optical loss. This optical loss causes a change in the output of the phototransistor, enabling the estimation of the contact force by measuring the phototransistor's output.

## 3 FABRICATION

Fig. 2 illustrates the fabrication process of the proposed tactile sensor. The core was made using a soft self-healing polymer (EMU6001, Yusiro Chemical Inc.). The contact part was fabricated by mixing black food dye (Food Dye Black, Kyoritsu Foods) into the self-healing polymer. First, a mold was printed using a 3D printer (Form3+, Formlabs) (Fig. 2(a)). Silicone rubber (Ecoflex00-50, Smooth-On) was then poured into the mold (Fig. 2(b)). After the silicone rubber cured, it was removed from the mold, and a diode (T-13 TSAL6100, Vishay Semiconductors) and a phototransistor (SFH 309 FA5/6, OSRAM Licht AG) were inserted along the grooves in the mold (Fig. 2(c)). Next, the self-healing polymer was poured into a larger silicone rubber mold and exposed to UV light with a wavelength of 405 nm for 30 seconds using a UV-LED curing light (Cure, SK Honpo) (Fig. 2(d)). For the smaller mold, the material was prepared by mixing 0.2% black food dye into EMU6001, and it was cured by exposing it to UV light with a wavelength of 405 nm for 300 seconds using a blacklight projector (UV Resin Curing Light, SUNPIE) (Fig. 2(d)). Finally, the components were removed from the silicone rubber mold, and the black contact area and the core were self-healed and connected (Fig. 2(e)). The fabricated sensor is shown in Fig. 2(f).

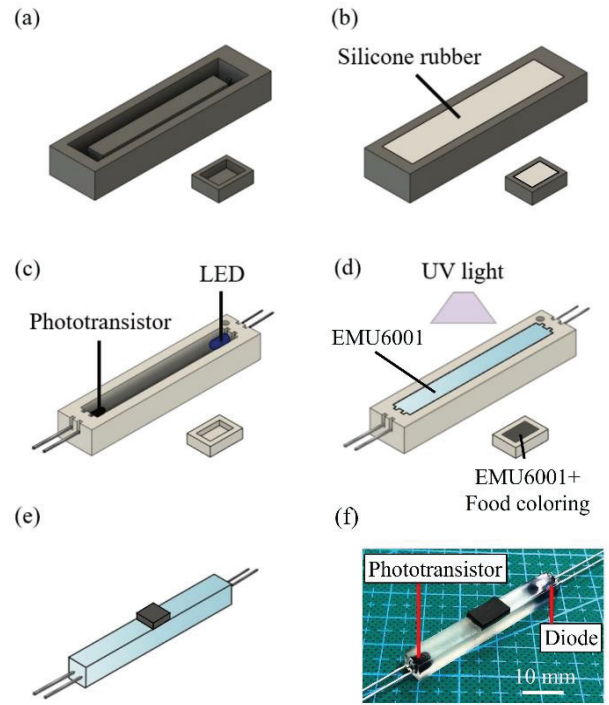

Fig.2 Fabrication process, (a) 3D printing the mold, (b) Injecting silicone rubber into the mold, (c) Inserting the diode and phototransistor, (d) Pouring self-healing material and UV curing, (e) Connecting the contact and optical waveguide, (f) Photo of the fabricated sensor.

## 4 EXPERIMENT

The response of the fabricated tactile sensor was evaluated when force was applied. The experimental setup used for the evaluation is shown in Fig. 3. The fabricated tactile sensor was fixed to a 3-axis table robot (ICSB3-BA1MB1L-WA-30AQ-25AQ-10AQB-T2-3L-CT-CT, IAI). Additionally, a 6-axis force sensor (PFS020YA500U6, Leptrino) with a square probe of 12 mm on each side was attached to the Z-axis slider of the 3-axis table robot, which was used to apply force to the tactile sensor by moving the probe. The probe was pressed at a speed of 1 mm/s, with a displacement of 3 mm. The output of the tactile sensor was recorded using a data recorder (NR-500, KEYENCE) at a sampling frequency of 1 kHz. Fig. 4 shows the results of the evaluation. The results confirmed that as the applied force increased, optical loss occurred, leading to a decrease in the sensor's output.

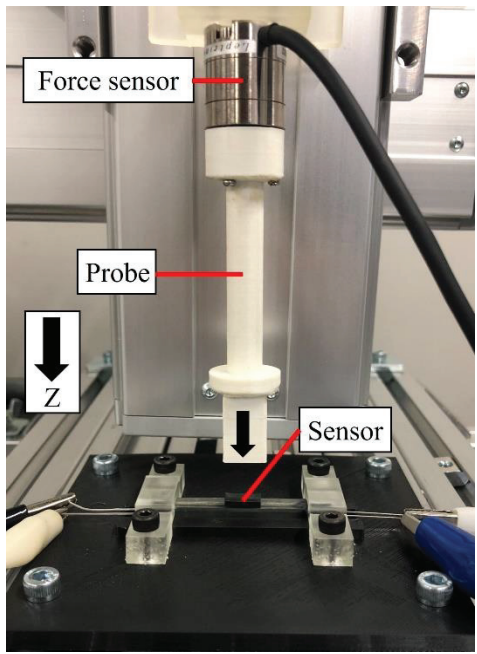

Fig.3 Experiment setup.

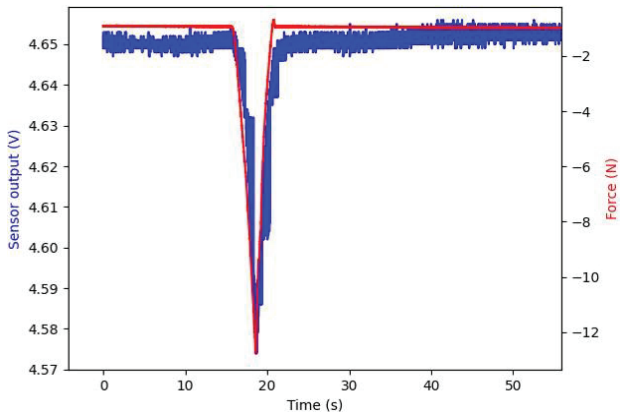

Fig.4 Results of the response for force input.

## 5 CONCLUSION

In this study, we developed an optical waveguide-based tactile sensor using a self-healing polymer. We determined the design requirements for an optical waveguide-based tactile sensor and successfully developed the fabrication process of the tactile sensor. The sensor's output was measured under applied load, and its characteristics were evaluated. The results confirmed that the sensor output decreased as the load increased. In the future, we plan to investigate the effect of self-healing on sensor characteristics before and after damage, as well as to mount the tactile sensor on a two-finger gripper for object identification.

ACKNOWLEDGEMENT

This work was supported in part by JSPS KAKENHI Grant Numbers JP22H01447 and JST PRESTO Grant Number JPMJPR22S2, Japan.